\documentclass{article}
\usepackage[dvips]{graphicx}  
\usepackage{color}

\usepackage[varg]{txfonts}
\usepackage{concmath}

\begin{document}

\newcommand{\rum}{\rule{0.5pt}{0pt}}
\newcommand{\rub}{\rule{1pt}{0pt}}
\newcommand{\rim}{\rule{0.3pt}{0pt}}
\newcommand{\numtimes}{\mbox{\raisebox{1.5pt}{${\scriptscriptstyle \rum\times}$}}}
\newcommand{\numtimess}{\mbox{\raisebox{1.0pt}{${\scriptscriptstyle \rum\times}$}}}
\newcommand{\Boldsq}{\vbox{\hrule height 0.7pt
\hbox{\vrule width 0.7pt \phantom{\footnotesize T}%
\vrule width 0.7pt}\hrule height 0.7pt}}
\newcommand{\two}{$\raise.5ex\hbox{$\scriptstyle 1$}\kern-.1em/
\kern-.15em\lower.25ex\hbox{$\scriptstyle 2$}$}

\renewcommand{\refname}{References}
\renewcommand{\tablename}{\small Table}
\renewcommand{\figurename}{\small Fig.}
\renewcommand{\contentsname}{Contents}

\twocolumn[%
\begin{center}
{\Large\bf 
Characterisation  of Low Frequency Gravitational Waves from    Dual RF Coaxial-Cable Detector: Fractal Textured Dynamical 3-Space\rule{0pt}{13pt}}\par

\bigskip
Reginald T. Cahill \\ 
\vspace{1mm}Progress in Physics   {\bf 3}, 3-10, 2012\\
{\small\it  School of Chemical and Physical  Sciences, Flinders University,
Adelaide 5001, Australia\rule{0pt}{15pt}}\\
\raisebox{+1pt}{\footnotesize E-mail: Reg.Cahill@flinders.edu.au}\par

\bigskip

{\small\parbox{11cm}{%
Experiments have revealed that the Fresnel drag effect is not present in RF coaxial cables, contrary to  a previous report. This enables a very sensitive, robust and compact detector, that is 1st order in v/c  and using one clock, to  detect the dynamical space passing the earth, revealing the sidereal rotation of the earth, together with significant wave/turbulence effects. These are ``gravitational waves", and previously detected  by   Cahill 2006,   using an  Optical-Fibre - RF Coaxial Cable Detector, and Cahill 2009, using a preliminary version of the Dual RF Coaxial Cable Detector. The gravitational waves have a 1/f spectrum, implying a fractal textured structure to dynamical 3-space. 
\rule[0pt]{0pt}{0pt}}}\medskip
\end{center}]

\setcounter{section}{0}
\setcounter{equation}{0}
\setcounter{figure}{0}
\setcounter{table}{0}

\markboth{Cahill R.T.  Characterisation  of Low Frequency Gravitational Waves from    Dual RF Coaxial-Cable Detector.}{\thepage}
\markright{Cahill R.T.  Characterisation  of Low Frequency Gravitational Waves from   Dual RF Coaxial-Cable Detector.}


\section{Introduction}
Data from a new   gravitational wave experiment is reported\footnote{This report is from the Gravitational Wave Detector Project at Flinders University.}, revealing a fractal textured 3-space, flowing past the earth at $\sim$500 km/s. The wave/turbulence or ``gravitational waves" have a significant magnitude, and are now known to have been detected numerous times over the last 125 years. The detector uses a single clock with RF EM waves propagating through dual coaxial cables, and is 1st order in $v/c$. The detector is sensitive, simple to operate, robust and compact.   It uses the surprising discovery that there is no Fresnel drag effect in coaxial cables, whereas there is in gases, optical fibres, liquids etc. Data from an analogous detector using optical fibres and single coaxial cables was reported in 2006 \cite{CahillGW,CahillBrisbane}. Because of the discovery reported herein that detector calibration has now been correctly redetermined.   Results from Michelson-Morley \cite{MMCK,MMC}, Miller \cite{Miller}, Torr and Kolen \cite{Torr} and  DeWitte \cite{DeWitte}, are now in remarkable agreement with the velocity of absolute motion of the earth determined from NASA spacecraft earth-flyby Doppler shift data \cite{And2008, CahillNASA}, all revealing a light/EM speed anisotropy of some 486km/s  in the direction RA=4.29$^h$, Dec=-75.0$^{\circ}$: that speed is $\sim$300,000-500 km/s  for radiation travelling in that direction, and $\sim$300,000+500 km/s travelling in the opposite, northerly,  direction: a significant observed anisotropy that physics has ignored. The actual daily average velocity varies with days of the year because of the orbital motion of the earth - the aberration effect discovered by Miller, but shows fluctuations  over all time scales.

In 2002 it was discovered that the Michelson-Morley 1887 light-speed anisotropy  experiment,  using the interferometer in gas mode,   had indeed detected anisotropy, by taking account of both the Lorentz length contraction effect  for the interferometer arms,  and the refractive index effect of the air in the light paths  \cite{MMCK,MMC}.     These gas-mode interferometer experiments show the difference between Lorentzian Relativity (LR)  and Special Relativity (SR).  In LR the length contraction effect is caused by motion of a rod, say,  through the dynamical 3-space, whereas in SR the length contraction is only a perspective effect, supposedly occurring only when the rod is moving relative to an observer.  This was further clarified  when an exact mapping between Galilean space   and time coordinates and the Minkowski-Einstein spacetime coordinates was recently discovered \cite{CahillMink}.  

 The Michelson interferometer, having  the calibration constant $k^2=(n^2-1)(n^2-2)$ where $n$ is the refractive index of the light-path medium, has zero sensitivity to EM anisotropy and gravitational waves when operated in vacuum-mode ($n=1$).  Fortunately the early experiments had air present in the light paths\footnote{Michelson and Morley implicitly assumed that $k^2=1$, which considerably overestimated the sensitivity of their detector by a factor of $\sim 1700$ (air has $n=1.00029$). This error lead to the invention of  ``spacetime" in 1905. Miller avoided any assumptions about the sensitivity of his detector, and used the earth orbit effect  to estimate the calibration factor $k^2$ from his data, although even that is now known to be incorrect: the sun 3-space inflow component was unknown to Miller. It was only in 2002 that the design flaw in the Michelson interferometer was finally understood  \cite{MMCK,MMC}.}. A very compact and cheap  Michelson interferometric anisotropy  and gravitational wave detector  may be constructed using optical fibres \cite{CahillOF},   but for most fibres $n\approx \sqrt 2$ near room temperature, and so needs to be operated at say 0$^\circ$C. The $(n^2-2)$  factor  is caused by the Fresnel drag\cite{Book}.
The detection of light speed anisotropy - revealing a flow of space past the detector,  is now entering an era of precision   measurements, as reported herein. These are  particularly important because experiments have shown large turbulence effects in the flow, and  are beginning to characterise this turbulence. Such turbulence can be shown to correspond to what are, conventionally, known  as gravitational waves, although not those implied by General Relativity, but more precisely are revealing a fractal structure to dynamical 3-space.

\section{Fresnel Drag}
The detection and characterisation of these wave/turbulence effects requires only the development  of small and cheap detectors, as these  effects are large.  However in all detectors the EM signals travel through a dielectric, either in bulk or optical fibre or through RF coaxial cables. For this reason it is important to understand the so-called Fresnel drag effect.  In optical fibres the Fresnel drag effect has been established, as this is important in the operation of  Sagnac optical fibre gyroscopes, for only then is the calibration independent of the fibre refractive index, as observed. The Fresnel drag effect is a phenomenological formalism that characterises the effect of the absolute motion of the propagation medium, say a dielectric,   upon the speed of the EM radiation relative to  that medium.

The Fresnel drag expression is that a dielectric in absolute motion through space at speed $v$, relative to space itself, causes the EM radiation to travel at speed
\begin{equation}
V(v)=\frac{c}{n}+v\left(1-\frac{1}{n^2}\right)
\label{eqn:Fresnel}\end{equation}
wrt the dielectric, when $V$ and $v$ have the same  direction. Here $n$ is the dielectric refractive index. The 2nd term is known as the Fresnel drag, appearing to show that the moving dielectric ``drags" the EM radiation, although   this is a misleading interpretation; see \cite{CahillFresnel} for a derivation\footnote{The Fresnel Drag in (\ref{eqn:Fresnel}) can be ``derived" from the Special Relativity velocity-addition formula, but there $v$ is the speed of the dielectric wrt to the observer, and as well common to both dielectrics and coaxial cables.}.   If the Fresnel drag is always applicable then, as shown herein, a 1st order in $v/c$ detector requires two clocks, though not necessarily synchronised, but requiring a rotation of the detector arm  to extract the $v$-dependent  term.  However, as we show herein, if the Fresnel drag is not present in RF coaxial cables, then  a detector 1st order in $v/c$  and using one clock, can detect and characterise  the dynamical  space. In \cite{CahillFresnel} it was incorrectly concluded that the Fresnel effect was present in RF coaxial cables, for reasons related to the temperature effects, and discussed later.

 \begin{figure}[t]
 \vspace{-7mm}
\hspace{5mm}\includegraphics[scale=1.0]{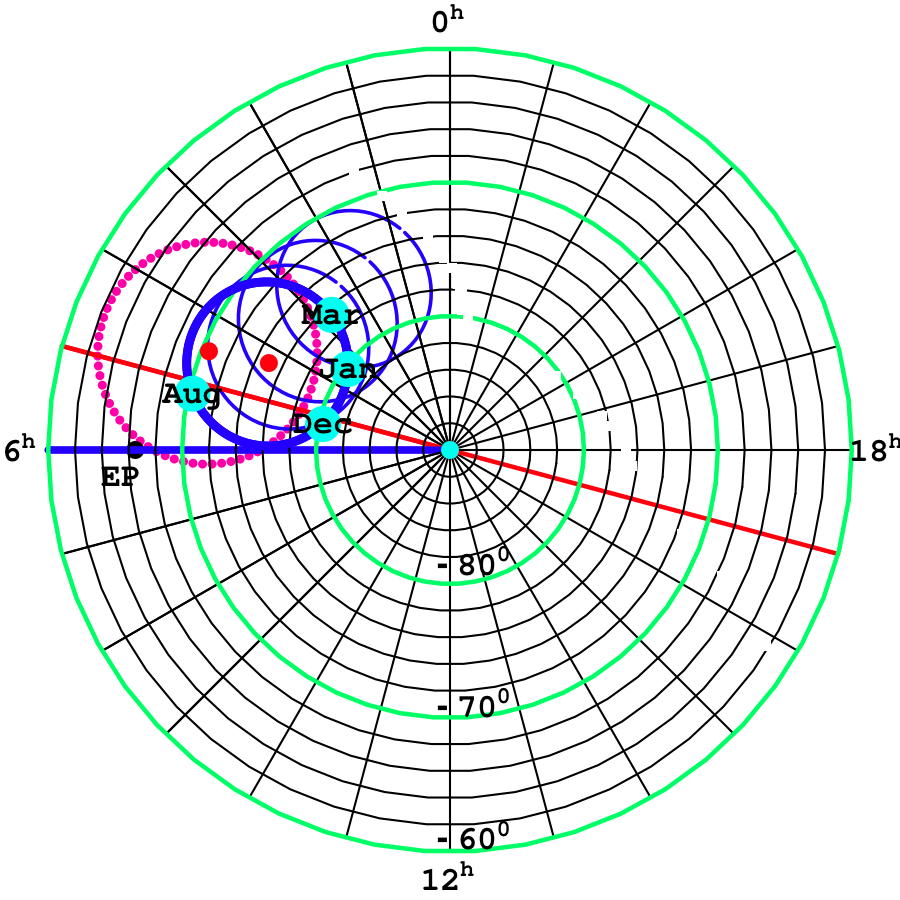}
\vspace{-5mm}	\caption{\small {South celestial pole region.  The dot (red)  at RA=4.29$^{h}$, Dec=-75$^\circ$, and with speed 486km/s, is the direction of motion of the solar system through space determined from NASA spacecraft earth-flyby Doppler shifts \cite{CahillNASA}, revealing  the EM radiation speed anisotropy. The thick (blue) circle centred on this direction is the observed velocity direction for different days of the year, caused by earth orbital motion and sun 3-space inflow. The corresponding results from Miller gas-mode interferometer are shown by 2nd dot (red) and its aberration circle (red dots) \cite{Miller}. For March the velocity is  RA=2.75$^{h}$, Dec=-76.6$^\circ$, and with speed 499.2km/s, see Table 2 of \cite{CahillNASA}. The thinner blue aberration  circles relate to determination of earth 3-space inflow speed, see \cite{CahillNASA}.}}
\label{fig:Aber}\end{figure}

\section{Dynamical 3-Space}
We briefly outline the dynamical modelling of 3-space. It involves the space velocity field ${\bf v}({\bf r},t)$, defined relative to an observer's frame of reference.
\begin{equation}
 \nabla\!\!\cdot\!\!\left(\!\!\frac{\partial\mathbf{v}}{\partial t}\!\!+\!\!\left(\mathbf{v}\!\cdot\!\nabla\!\right)\mathbf{v}\!\!\right)\!+\!
 \frac{\alpha}{8}\!\!\left(\!\left(\mathrm{tr}D\right)^{2}\!-\!\mathrm{tr}\!\left(D\right)^{2}\right)\!\! +..=-4\pi G\rho
\label{eqn:3spaceequation}
\end{equation}
$\nabla \times \mathbf{v}=\mathbf{0}$ and  $D_{ij}=\partial v_i/\partial x_j$. The velocity field ${\bf v}$  describes classically the time evolution of the substratum quantum foam. The bore hole $g$ anomaly data has revealed $\alpha = 1/137$,  the fine structure constant.
The matter acceleration is found by determining the trajectory of a quantum matter wavepacket. This is most easily done by maximising the proper travel time $\tau$:
 \begin{equation}
\tau=\int dt \sqrt{1-\frac{{\bf v}^2_R({\bf r}_0(t),t)}{c^2}}
\label{eqn:propertime}\end{equation} 
where ${\bf v}_R({\bf r}_o(t),t) ={\bf v}_o(t) - {\bf v}({\bf r}_o(t),t),$ is the velocity of the wave packet, at position ${\bf r}_0(t)$,  wrt the local space - a neo-Lorentzian Relativity effect.
 This ensures that quantum waves propagating along neighbouring paths are in phase, and so interfere constructively. 
This maximisation gives the quantum matter geodesic equation for ${\bf r}_0(t)$
\begin{equation}
{\bf g}=\displaystyle{\frac{\partial {\bf v}}{\partial t}}+({\bf v}\cdot{\bf \nabla}){\bf
v}+({\bf \nabla}\times{\bf v})\times{\bf v}_R-\frac{{\bf
v}_R}{1-\displaystyle{\frac{{\bf v}_R^2}{c^2}}}
\frac{1}{2}\frac{d}{dt}\left(\frac{{\bf v}_R^2}{c^2}\right)+...
\label{eqn:acceleration}\end{equation}  
with  ${\bf g}\equiv d{\bf v}_o/dt=d^2{\bf r}_o/dt^2$.  
The 1st term in $\bf g$ is  the Euler space acceleration   $\bf a$, the 2nd term explains the Lense-Thirring effect, when the vorticity is non-zero,  and the last term   explains the precession of orbits.
While the velocity field has been repeatedly detected since the Michelson-Morley 1887 experiment, the best detection has been using the spacecraft earth-flyby Doppler shift data \cite{CahillNASA},
see Fig.\ref{fig:Aber}. The above reveals gravity to be an emergent phenomenon where quantum matter waves are refracted by the time dependent and inhomogeneous 3-space velocity field. The $\alpha$-term in (\ref{eqn:3spaceequation}) explains the so-called ``dark matter" effects:  if $\alpha\rightarrow 0$ and $v_R/c\rightarrow 0$ we derive Newtonian gravity, for then $\nabla\cdot{\bf g}=-4\pi G \rho$ \cite{Book}. Note that the relativistic term in (\ref{eqn:acceleration}) arises from the quantum matter dynamics - not from the space dynamics.

\section{Gravitational Waves: Dynamical Fractal 3-Space}
Eqn.(\ref{eqn:propertime}) for the elapsed proper time maybe written in differential form as
\begin{equation}
d\tau^2\!\!=\!dt^2\!\!-\frac{1}{c^2}(d{\bf r}(t)\!-\!{\bf v}({\bf r}(t),t)dt)^2\!\!=\!\!g_{\mu\nu}(x)dx^\mu dx^\nu
\label{eqn:PGmetric}\end{equation}
which introduces a curved spacetime metric $g_{\mu\nu}$ for which the geodesics are the quantum matter trajectories when freely propagating through the dynamical 3-space.  Gravitational wave  are traditionally thought of as ``ripples" in the space-time metric $g_{\mu\nu}$. But the discovery of the dynamical 3-space means that they are more appropriately understood to be turbulence effects of the dynamical 3-space vector $\bf v$, because it is ${\bf v}$ that is directly detectable, whereas $g_{\mu\nu}$ is merely an induced mathematical artefact. When the matter density $\rho=0$,  (\ref{eqn:3spaceequation}) will have a time-dependent  fractal structure solutions, as there is no length scale.   The wave/turbulence effects reported herein confirm that prediction, see Fig.\ref{fig:Space}.

 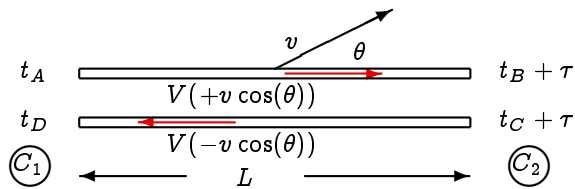
\begin{figure}
\vspace{10mm}
\hspace{-5mm}
\setlength{\unitlength}{1.3mm}
\hspace{0mm}\begin{picture}(0,0)
\thicklines

\definecolor{hellgrau}{gray}{.8}
\definecolor{dunkelblau}{rgb}{0, 0, .9}
\definecolor{roetlich}{rgb}{1, .7, .7}
\definecolor{dunkelmagenta}{rgb}{.9, 0, .0}
\definecolor{green}{rgb}{0, 1,0.4}
\definecolor{black}{rgb}{0, 0, 0}

\color{dunkelmagenta}
\put(30,5.5){\vector(1,0){10}}
\put(25,0.5){\vector(-1,0){10}}

\color{black}

\put(9,5){\line(1,0){40}}
\put(9,6){\line(1,0){40}}

\put(9,5){\line(0,1){1}}
\put(49,5){\line(0,1){1}}

\put(9,0){\line(1,0){40}}
\put(9,1){\line(1,0){40}}

\put(9,0){\line(0,1){1}}
\put(49,0){\line(0,1){1}}

\put(3,5){{$t_A$}}
\put(3,0){{$t_D$}}

\put(52,5){{$t_B+\tau$}}
\put(52,0){{$t_C+\tau$}}

\put(29,6){\vector(2,1){12}}
\put(30,8){{$v$}}
\put(37,6.9){{$\theta$}}

\put(18,2.5){{$V(+v\cos(\theta))$}}
\put(18,-2.5){{$V(-v\cos(\theta))$}}

\put(4,-4){\circle{4}}
\put(2.2,-4.5){{$C_1$}}

\put(55,-4){\circle{4}}
\put(53.3,-4.5){{$C_2$}}

\put(20,-5){\vector(-1,0){11}}
\put(25,-6){$ L$}
\put(32,-5){\vector(1,0){17}}

\end{picture}

\vspace{8mm}
	\caption{\small{ Schematic layout for measuring the one-way speed of light in either free-space, optical fibres or RF coaxial cables, without requiring the synchronisation of the clocks $C_1$ and $C_2$: here $\tau$ is the unknown offset time between the clocks.  $V$ is the speed of EM radiation wrt the apparatus, with or without the Fresnel drag in (\ref{eqn:Fresnel}), and $v$ is the speed of the apparatus through space, in direction $\theta$. \mbox{DeWitte} used this technique in 1991 with1.5km RF cables and Cesium atomic clocks \cite{DeWitte}. In comparison with data from spacecraft earth-flyby Doppler shifts \cite{CahillNASA} this experiments confirms that there is no Fresnel drag effect in RF coaxial cables.}}
 \label{fig:oneway}
\end{figure}

 \begin{figure}[t]
 \vspace{-3mm}
 \hspace{3.5mm}\includegraphics[scale=0.525]{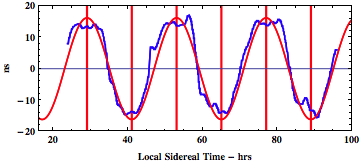}
 
 \hspace{-3mm}\includegraphics[scale=0.22]{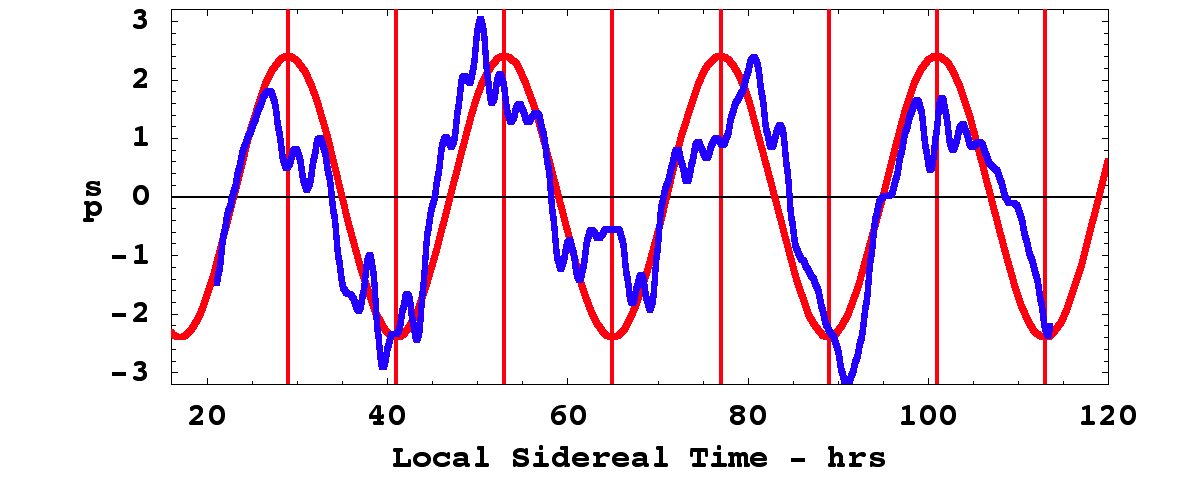}
\vspace{-7mm}	\caption{\small {Top:  Data from the 1991 DeWitte NS horizontal coaxial cable experiment, $L=1.5$km,  $n=1.5$, using the arrangement shown in Fig.\ref{fig:oneway}. The time variation of $\sim 28$ns  is consistent with the Doppler shift results   with speed 500km/s, but using Dec=-65$^\circ$: the month for this data is unknown, and the vertical red lines are at RA=5$^{h}$.  If a Fresnel drag effect is included the speed would have to be 1125km/s, in disagreement with the Doppler shift data, demonstrating that there is no Fresnel drag in coaxial cables.  Bottom:   Dual coaxial cable detector  data  from May  2009 using the technique in Fig.\ref{fig:DualCoax} and without looping: $L=20$m, Doppler shift data predicts Dec$=-77^\circ$, $v=480$km/s giving a sidereal dynamic range of 5.06ps, very close to that observed.  The  vertical red lines are at RA=5$^{h}$. In both data sets we see the earth sidereal rotation effect together with significant wave/turbulence effects.}}
\label{fig:Coax2009}\end{figure}

\section{First Order in $\bf v/c$ Speed of EMR Experiments}
Fig.\ref{fig:oneway} shows the arrangement for measuring the one-way speed of light, either in vacuum, a dielectric, or RF coaxial cable.  It is usually argued that one-way speed of light measurements are not possible because the clocks $C_1$ and $C_2$ cannot be synchronised.   However  this is false, and at the same time shows an important  consequence of  
(\ref{eqn:Fresnel}). In the upper part of  Fig.\ref{fig:oneway} the actual travel time $t_{AB}$ from $A$ to $B$ is determined by
\begin{equation}
V(v\cos(\theta))t_{AB}=L+v\cos(\theta)t_{AB}
\label{eqn:traveltime1}\end{equation}
where the 2nd term comes from the end $B$ moving an additional distance  $v\cos(\theta)t_{AB}$ during time interval $t_{AB}$.  Then
\begin{equation}
t_{AB}= \frac{L}{V(v\cos(\theta))-v\cos(\theta)}=\frac{nL}{c}+\frac{v\cos(\theta)L}{c^2}+..
\label{eqn:traveltime2}\end{equation}
\begin{equation}
t_{CD}= \frac{L}{V(v\cos(\theta))+v\cos(\theta)}=\frac{nL}{c}-\frac{v\cos(\theta)L}{c^2}+..
\label{eqn:traveltime3}\end{equation}
on using (\ref{eqn:Fresnel}), i.e. assuming the validity of the Fresnel effect,  and expanding to 1st oder in $v/c$.  However if there is no Fresnel drag effect then we obtain 
\begin{equation}
t_{AB}\!=\! \frac{L}{V(v\cos(\theta))-v\cos(\theta)}\!=\!\frac{nL}{c}\!+\frac{v\cos(\theta)Ln^2}{c^2}+..
\label{eqn:traveltime2b}\end{equation}
\begin{equation}
t_{CD}\!=\! \frac{L}{V(v\cos(\theta))\!+\!\!v\cos(\theta)}\!=\!\frac{nL}{c}-\!\frac{v\cos(\theta)Ln^2}{c^2}+..
\label{eqn:traveltime3b}\end{equation}
The  important observation is that the $v/c$ terms are independent of the dielectric refractive index $n$ in (\ref{eqn:traveltime2}) and  (\ref{eqn:traveltime3}), but have an $n^2$ dependence in (\ref{eqn:traveltime2b}) and  (\ref{eqn:traveltime3b}), in the absence of the Fresnel drag effect.  

 If the clocks are not synchronised then  $t_{AB}$ is not known, but by changing direction of the light path, that is varying $\theta$, the magnitude of the 2nd term may be separated from the magnitude of the 1st term, and $v$ and its direction determined.  The clocks may then  be synchronised.   For a small detector the change in $\theta$ can be achieved by a  direct rotation.  Results (\ref{eqn:traveltime2}) and  (\ref{eqn:traveltime3}), or (\ref{eqn:traveltime2b}) and  (\ref{eqn:traveltime3b}),  have been exploited in various detector designs.

\section{DeWitte 1st Order in $\bf v/c$ Detector}
 The DeWitte  $L=1.5$km RF coaxial cable experiment, Brussels 1991,   was a double 1st order in $v/c$ detector, using the scheme in Fig.\ref{fig:oneway}, and  employing  3 Caesium  atomic clocks at each end, and overall measuring $t_{AB}-t_{CD}$.  The orientation was NS and rotation was achieved by that of the earth \cite{DeWitte}.   
 \begin{equation}
 t_{AB}-t_{CD}=\frac{2v\cos(\theta)Ln^2}{c^2}
 \label{eqn:DeWitte}\end{equation}
 The dynamic range of $\cos(\theta)$ is $2\sin(\lambda-\beta)\cos(\delta)$, caused by the earth rotation, where $\lambda$ is the latitude of the detector location, $\beta$ is the local inclination to the horizontal, here $\beta=0$, and $\delta$ is the declination of $\bf v$. The data shows remarkable agreement with the velocity vector from the flyby Doppler shift data, see Fig.\ref{fig:Coax2009}. However if there is Fresnel drag in the coaxial cables, there would be  no $n^2$ factor in (\ref{eqn:DeWitte}), and the DeWitte data would give a much larger speed $v=1125$km/s, in strong disagreement with the flyby data.
 
 \section{Torr and Kolen 1st Order in $\bf v/c$ Detector}
   \begin{figure}
\vspace{0mm}
\hspace{4mm}\includegraphics[scale=1.0]{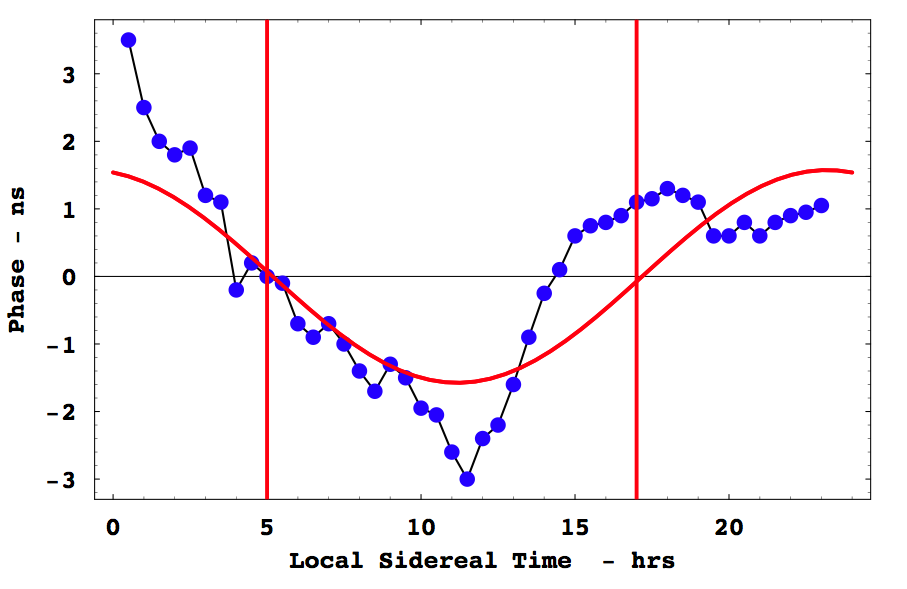}
\vspace{-2mm}
\caption{\small{Data from the 1981 Torr-Kolen experiment at Logan, Utah \cite{Torr}. 
The data shows variations in travel times
(ns),  for local sidereal times,  of an RF signal travelling through $500$ m of coaxial 
cable  orientated in an EW direction.  The red curve is sidereal effect prediction  for   February, 
for a 3-space speed of $480$ km/s from the direction, RA=$5^h$, Dec=-$70^\circ$.  }}
\label{fig:TorrKolen}\end{figure}

A one-way coaxial cable\index{RF coaxial cable!Torr-Kolen} experiment  was
performed at the Utah University in 1981 by Torr and  Kolen \cite{Torr}. This involved two Rb clocks placed approximately $500$ m apart with a 5 MHz sinewave RF signal propagating between the clocks via a nitrogen filled coaxial cable buried in the ground and maintained at a constant pressure of $\sim$2 psi. Torr and Kolen  observed  variations in the one-way travel time, as shown in Fig.\ref{fig:TorrKolen} by the  data points.   The theoretical  predictions for the  Torr-Kolen experiment for a cosmic speed of $480$ km/s from the direction,  RA=$5^h$, Dec=-$70^\circ$, as  shown in Fig.\ref{fig:TorrKolen}.   The maximum/minimum effects occurred, typically,  at the predicted times.    Torr and Kolen reported fluctuations in both the magnitude, from 1 - 3 ns, and time of the maximum variations in travel time, just as  observed in all  later experiments, namely wave effects.

 \begin{figure*}[t]
   \hspace{6mm}\includegraphics[scale=0.3]{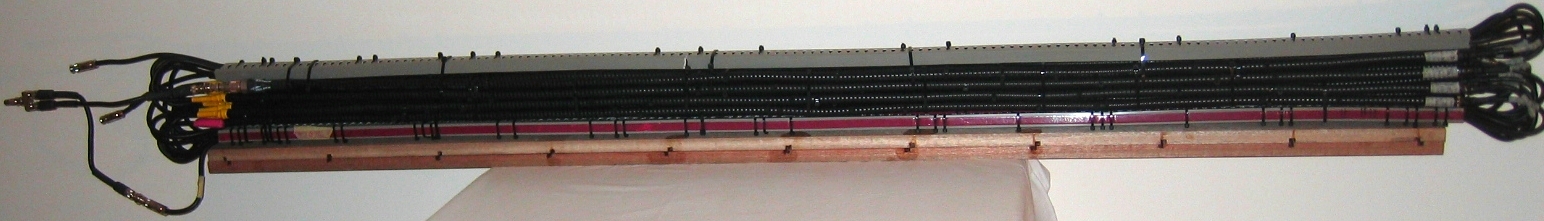}  
   \vspace{-2mm}
\caption{\small{Photograph of the RF coaxial cables arrangement, based upon 16 $\times$1.85m lengths of phase stabilised  Andrew HJ4-50 coaxial cable. These are joined to 16 lengths of phase stabilised  Andrew  FSJ1-50A cable, in the manner shown schematically in Fig.\ref{fig:DualCoax}.  The 16 HJ4-50 coaxial cables have been tightly bound into a 4$\times$4 array, so that the cables, locally, have the same temperature, with cables in one of the circuits embedded between cables in the 2nd circuit.  This arrangement of the cables permits the cancellation of temperature differential effects in the cables. A similar array of the smaller  diameter FSJ1-50A cables  is located inside the grey-coloured conduit  boxes.   }}
\label{fig:coaxphoto}\end{figure*}

   \begin{figure}[t]
\vspace{5mm}
\setlength{\unitlength}{1.1mm}
\setlength{\unitlength}{1.1mm}

\vspace{12mm}

\hspace{4mm}\begin{picture}(0,0)
\thicklines

\definecolor{hellgrau}{gray}{.8}
\definecolor{dunkelblau}{rgb}{0, 0, .9}
\definecolor{roetlich}{rgb}{1, .7, .7}
\definecolor{dunkelmagenta}{rgb}{.9, 0, .0}
\definecolor{mauve}{rgb}{0.4, 0, .8}

\put(0,6.5){\bf S}\put(65,6.5){\bf N}

\put(5.5,-2){\bf A}
\put(67,-2){\bf B}

\put(5.5,16){\bf D}
\put(67,16){\bf C}

\put(0,0){\line(0,1){5}}\put(0,0){\line(1,0){5}}
\put(0,5){\line(1,0){5}}\put(5,0){\line(0,1){5}}
\put(0.2,1.9){\bf Rb}

\put(0,10){\line(0,1){5}}\put(0,10){\line(1,0){5}}
\put(0,15){\line(1,0){5}}\put(5,10){\line(0,1){5}}
\put(0.4,12.6){\bf DS}
\put(0.6,10.2){\bf O}

\put(20,-5){\vector(-1,0){15}}
\put(25,-5.5){\large$ L$}
\put(32,-5){\vector(1,0){37}}

 \color{dunkelmagenta}
  \put(32,-2){\bf FSJ1-50A}
  \put(32,8.0){\bf FSJ1-50A}
\put(5,1.1){\line(1,0){64}}
\put(35,1.21){\vector(1,0){1}}
\put(35,11.3){\vector(-1,0){1}}
 
\color{dunkelblau}
  \put(32,15){\bf HJ4-50}
    \put(32,5){\bf HJ4-50}
\put(5,4){\line(1,0){64}}
\put(5,3.6){\line(1,0){64}}

  \color{dunkelmagenta}
\put(5,11.4){\line(1,0){64}}

 \color{dunkelblau}
\put(5,14){\line(1,0){64}}
\put(5,13.6){\line(1,0){64}}

\put(35,13.8){\vector(-1,0){1}}

\put(35,3.9){\vector(1,0){1}}

 \color{dunkelmagenta}
\put(68.4,8.92){\oval(5,10.1)[rb]}
\put(68.4,6.3){\oval(5,10.1)[rt]}

\put(68.5,6.2){\oval(10,10.1)[rb]}
\put(68.5,8.8){\oval(10,10)[rt]}
\put(73.5,6){\line(0,1){4}}
\put(73.6,6){\vector(0,1){3}}
\put(70.9,6.0){\vector(0,1){2}}

\end{picture}

\vspace{5mm}\caption{\small Because Fresnel drag is absent in RF coaxial cables this dual cable setup, using one clock, is capable of detecting the absolute motion of the detector  wrt to space, revealing the sidereal rotation effect as well as wave/turbulence effects. In the 1st trial of this detector this arrangement was used, with the cables laid out on a laboratory floor, and preliminary results are shown in Figs.\ref{fig:Coax2009}. In the new design  the cables in each circuit are configured into 8 loops, as in Fig.\ref{fig:coaxphoto}, giving $L=8\times1.85\mbox{m}=14.8$m. In \cite{CahillGW} a version with optical fibres in place of  the HJ4-50 coaxial cables was used, see Fig\ref{fig:FUGWD}. There the optical fibre has a Fresnel drag effect while the coaxial cable did not. In that experiment optical-electrical converters were used to modulate/demodulate infrared light.}
\label{fig:DualCoax}
\end{figure}
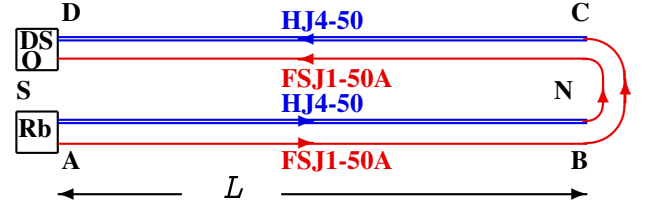

\section{Dual RF Coaxial Cable Detector}
The Dual RF Coaxial Cable Detector exploits the Fresnel drag anomaly,  in that there is no Fresnel drag effect in RF coaxial cables.  Then from  (\ref{eqn:traveltime2b}) and  (\ref{eqn:traveltime3b})  the round trip  travel time is, see  Fig.\ref{fig:DualCoax},
\vspace{-4mm}
\begin{equation}
t_{AB}+t_{CD}=\frac{(n_1+n_2)L}{c}+\frac{v\cos(\theta)L(n_1^2-n_2^2)}{c^2}+..
\label{eqn:RF1}\end{equation}
where $n_1$ and $n_2$ are the  effective refractive indices for the two  different RF coaxial cables, with two separate circuits to reduce temperature effects.  Shown in Fig.\ref{fig:coaxphoto} is a photograph.  The Andrews Phase Stabilised FSJ1-50A has $n_1=1.19$,  while  the HJ4-50 has $n_2 = 1.11$. One measures the travel time difference of  two RF signals from a Rubidium frequency standard (Rb) with a Digital Storage Oscilloscope (DSO). In each circuit the RF signal travels one-way in one type of coaxial cable, and returns via a different kind of coaxial cable.  Two circuits are used so that temperature effects cancel - if a temperature change alters the speed in one type of cable, and so the travel time, that travel time change is the same in both circuits, and cancels in the difference.  The travel time difference of the two circuits at the DSO is
\vspace{-2mm}
 \begin{equation}
\Delta t=\frac{2v\cos(\theta)L(n_1^2-n_2^2)}{c^2}+..
\label{eqn:RF2}\end{equation}
If  the Fresnel drag effect occurred in RF coaxial cables, we would use  (\ref{eqn:traveltime2}) and  (\ref{eqn:traveltime3}) instead, and then the  $n_1^2-n_2^2$ term is replaced by $0$, i.e. there is no 1st order term in $v$. That is contrary to the actual data in Figs.\ref{fig:Coax2009} and \ref{fig:WaveSpectrum}.

 The preliminary layout for this detector used cables laid out as in Fig.\ref{fig:DualCoax}, and the data is shown in Fig.\ref{fig:Coax2009}. In the compact  design the Andrew HJ4-50 cables are cut into 8 $\times$ 1.85m shorter lengths in each circuit, corresponding to a net length of $L=8\times 1.85 =14.8$m, and the Andrew FSJ1-50A cables are also cut, but into longer lengths to enable joining. However  the curved parts of the Andrew FSJ1-50A cables contribute only at 2nd order in $v/c$.   
The apparatus was horizontal, $\beta=0$, and orientated NS,  and used the rotation of the earth to change the angle $\theta$.   The dynamic range of $\cos(\theta)$, caused by the earth rotation only, is  again  $2\sin(\lambda-\beta)\cos(\delta)$,  where $\lambda=-35^\circ$ is the latitude of Adelaide.   Inclining the detector at angle $\beta=\lambda$  removes the earth rotation effect, as now the detector arm is parallel to the earth's spin axis, permitting  a more accurate characterisation of the wave effects.

\section{Temperature Effects}
The cable travel times  and the DSO phase measurements are  temperature dependent, and these effects are removed from the data, rather than attempt to maintain a constant temperature, which is impractical because of the heat output of the Rb clock and DSO. The detector was located in a closed room in which the temperature changed slowly over many days, with variations originating from changing external weather driven temperature changes. The temperature of the detector  was measured, and it was assumed that the timing errors were proportional to changes in that one measured temperature.  These timing errors were some 30ps, compared to the true signal of some 8ps.  Because the temperature timing errors are much larger,  the  temperature induced
$\Delta t= a+b\Delta T$ was fitted to the timing data, and the coefficients $a$ and $b$ determined.  Then this $\Delta t$ time series was subtracted from the data, leaving the actual required phase data.  This is particularly effective as the temperature variations had a distinctive time signature. The resulting data is shown in Fig.\ref{fig:Data}. In an earlier test for the Fresnel drag effect in RF coaxial cables  \cite{CahillFresnel} 
the technique for removing the temperature induced timing errors was inadequate, resulting in the wrong conclusion that there was Fresnel drag in RF coaxial cables.

\section{Dual RF Coaxial Cable Detector: Data}
\begin{figure}
\hspace{12mm}\includegraphics[scale=0.55]{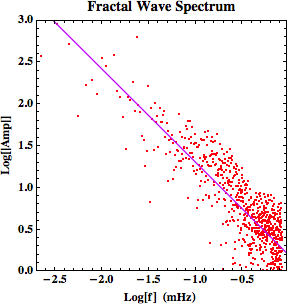}
	\caption{\small {Log-Log plot of the data  (top) in Fig.\ref{fig:WaveSpectrum}, with the straight line being $A\propto 1/f$, indicating a $1/f$ fractal wave spectrum.  The interpretation for this is the 3-space structure shown in Fig.\ref{fig:Space}.}}
\label{fig:WaveSpectrum}\end{figure}

\begin{figure*}[t]  
  \hspace{21mm}\includegraphics[scale=0.83]{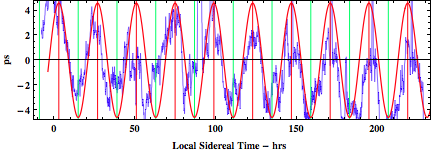}  
  
    \vspace{3mm}
   \hspace{22mm}\includegraphics[scale=0.8 ]{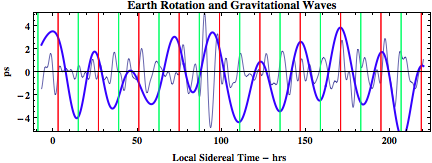}  
   \vspace{-2mm}
\caption{\small{Top: Travel time differences (ps) between the two coaxial cable circuits in Fig.\ref{fig:DualCoax}, orientated NS and horizontal,  over 9 days (March 4-12, 2012, Adelaide) plotted against local sidereal time.  Sinewave, with dynamic range 8.03ps,  is  prediction for  sidereal effect   from flyby Doppler shift data for RA=2.75$^h$ (shown by red fudicial lines), Dec=-76.6$^\circ$,  and with speed 499.2km/s, see Table 2 of \cite{CahillNASA},  also shown in from Fig.\ref{fig:Aber}. Data shows sidereal effect and significant wave/turbulence effects.  Bottom: Data filtered into two frequency bands $3.4\times 10^{-3} \mbox{mHz} <f < 0.018\mbox{mHz}$  ($81.4 h >T >15.3h$) and $0.018\mbox{mHz} <f < 0.067\mbox{mHz}$ ($15.3h >T>4.14h$), showing more clearly the earth rotation sidereal effect (plus vlf waves) and the turbulence without the sidereal effect.  Frequency spectrum of top data is shown in Fig.\ref{fig:WaveSpectrum}.}}
\label{fig:Data}
\end{figure*}

\begin{figure}
\hspace{1mm}\includegraphics[scale=0.35]{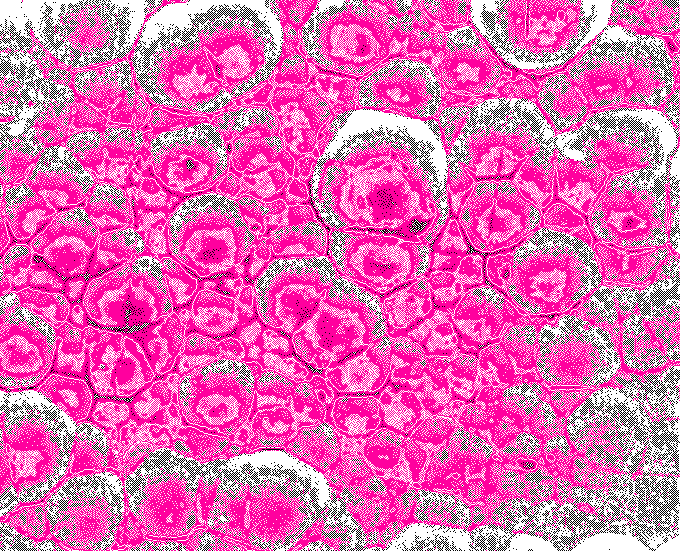}
	\caption{\small {Representation of the fractal wave data as a  revealing the fractal textured structure of the 3-space, with cells of space having slightly different velocities, and continually changing, and moving wrt the earth with a speed of $\sim$500km/s.   }}
\label{fig:Space}\end{figure}

 The phase data, after removing the temperature effects, is shown in Fig.\ref{fig:Data} (top), with the data compared with predictions  for the sidereal effect only from the flyby Doppler shift data. As well that data is separated into two frequency bands (bottom), so that the sidereal effect is partially separated from the gravitational wave effect, {\it viz} 3-space wave/turbulence. Being 1st order in $v/c$ it is easily determined that the space flow is from the southerly direction, as also reported in \cite{CahillGW}.  Miller reported the same sense, i.e. the flow is essentially from S to N, though using a 2nd order detector that is more difficult to determine.  The frequency spectrum of this data is shown in Fig.\ref{fig:WaveSpectrum}, revealing a fractal $1/f$ form.  This implies the fractal structure of the 3-space indicated in Fig.\ref{fig:Space}.
 
\begin{figure}[t]
\hspace{-2mm}\includegraphics[scale=0.27]{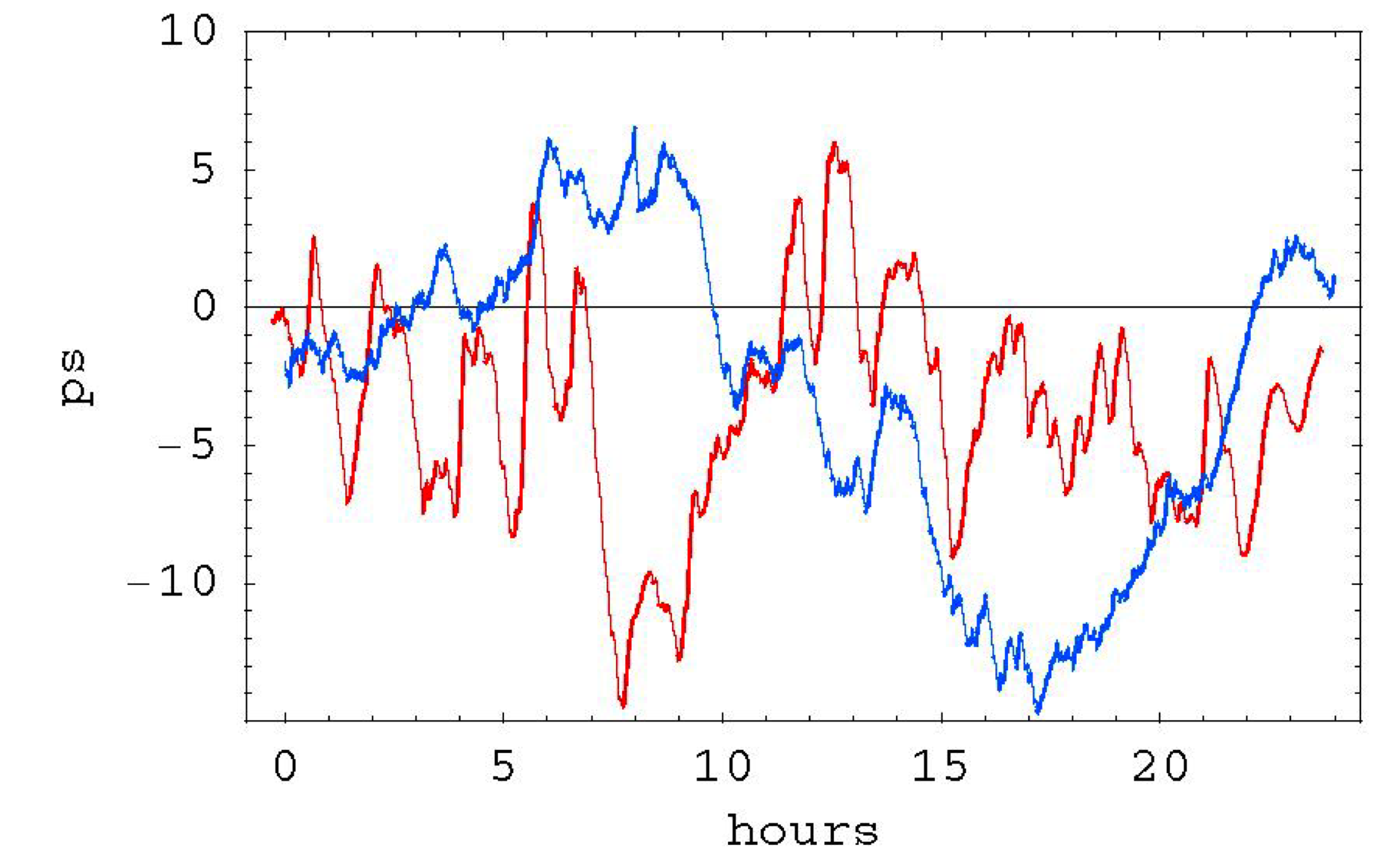}

\vspace{-1.5mm}

\caption{\small{ Phase difference (ps), with arbitrary zero,  versus local time data plots from the Optical Fibre - Coaxial Cable Detector, see Fig.\ref{fig:FUGWD}  and \cite{CahillGW,CahillBrisbane}, showing the sidereal time effect and significant wave/turbulence effects.. The plot (blue)  with the most easily identified minimum at $\sim$17\,hrs local Adelaide time is from June 9, 2006, while the plot  (red) with the minimum at $\sim$8.5hrs local time is from August 23, 2006. We see that the minimum has moved forward in time by approximately 8.5\,hrs.  The expected sidereal shift for this 65 day difference, without wave effects, is 4.3\,hrs,  to which must be added another $\sim$1h from the aberration effects shown in Fig\ref{fig:Aber}, giving 5.3hrs, in agreement with the data, considering that on individual days the min/max fluctuates by $\pm2$hrs.  This sidereal time shift is a critical test for the  detector.  From the flyby Doppler data we have for  August  RA=5$^h$, Dec=-70$^\circ$,  and  speed 478km/s,   see Table 2 of \cite{CahillNASA},  the predicted sidereal effect dynamic range to be 8.6ps, very close to that observed.}}
\label{fig:STEffect}
\end{figure}

\section{Optical Fibre RF Coaxial Cable Detector}
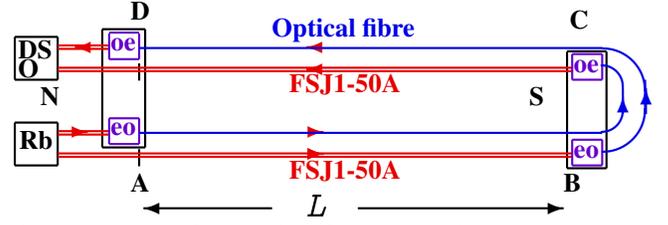
\begin{figure}
\setlength{\unitlength}{1.14mm}
\vspace{20mm}
\hspace{1mm}\begin{picture}(0,0)
\thicklines
\definecolor{hellgrau}{gray}{.8}
\definecolor{dunkelblau}{rgb}{0, 0, .9}
\definecolor{roetlich}{rgb}{1, .7, .7}
\definecolor{dunkelmagenta}{rgb}{.9, 0, .0}
\definecolor{mauve}{rgb}{0.4, 0, .8}

\put(30,-5){\vector(-1,0){15}}
\put(34,-6){\large$ L$}
\put(40,-5){\vector(1,0){24}}

\put(13.5,-3){\bf A}
\put(14.5,-0.1){\line(0,1)2}
\put(64.0,-3){\bf B}

\put(13.5,17){\bf D}
\put(14.5,9.9){\line(0,1)2}
\put(64.7,16){\bf C}

\put(0,0){\line(0,1){5}}\put(0,0){\line(1,0){5}}
\put(0,5){\line(1,0){5}}\put(5,0){\line(0,1){5}}
\put(0.5,1.9){\bf Rb}

\put(0,10){\line(0,1){5}}\put(0,10){\line(1,0){5}}
\put(0,15){\line(1,0){5}}\put(5,10){\line(0,1){5}}
\put(0.4,12.5){\bf DS}
\put(0.4,10.25){\bf O}

\put(10.2,2.1){\line(0,1){13.8}}
\put(10.2,2.1){\line(1,0){5.0}}
\put(10.2,15.9){\line(1,0){5.0}}
\put(15.2,2.1){\line(0,1){13.8}}

\put(64.4,-0.33){\line(0,1){13.6}}
\put(64.4,-0.33){\line(1,0){4.6}}
\put(64.4,13.32){\line(1,0){4.6}}
\put(69.0,-0.33){\line(0,1){13.6}}

\put(3,7){\bf N}\put(60,7){\bf S}
  \color{dunkelmagenta}
\put(5,1){\line(1,0){60}}
\put(5,1.4){\line(1,0){60}}
\put(35,1.35){\vector(1,0){1}}

\put(5,4){\line(1,0){6}}
\put(5,3.6){\line(1,0){6}}

\put(5,11){\line(1,0){60}}
\put(5,11.4){\line(1,0){60}}

\put(5,14){\line(1,0){6}}
\put(5,13.6){\line(1,0){6}}

\put(35,1.35){\vector(1,0){1}}
\put(35,13.8){\vector(-1,0){1}}
\put(8,13.8){\vector(-1,0){1}}
\put(35,11.2){\vector(-1,0){1}}
\put(35,3.9){\vector(1,0){1}}

\put(7.5,3.9){\vector(1,0){1}}

  \put(32,8.5){\bf FSJ1-50A}
  \put(32,-1.5){\bf FSJ1-50A}

\color{mauve}

\put(11,2.5){\line(0,1){3}}
\put(11,2.5){\line(1,0){3.5}}
\put(11,5.5){\line(1,0){3.5}}
\put(14.5,2.5){\line(0,1){3}}
\put(11.2,3.5){\bf eo}

\put(65,0.0){\line(0,1){3}}
\put(65,0.0){\line(1,0){3.5}}
\put(65,3.0){\line(1,0){3.5}}
\put(68.5,0){\line(0,1){3}}
\put(65.2,0.9){\bf eo}

\put(65,10.0){\line(0,1){3}}
\put(65,10.0){\line(1,0){3.5}}
\put(65,13.0){\line(1,0){3.5}}
\put(68.5,10){\line(0,1){3}}
\put(65.2,10.9){\bf oe}

\put(11,12.4){\line(0,1){3}}
\put(11,12.4){\line(1,0){3.5}}
\put(11,15.5){\line(1,0){3.5}}
\put(14.5,12.4){\line(0,1){3}}
\put(11.2,13.4){\bf oe}

\color{dunkelblau}
   \put(30,15.0){\bf Optical fibre}
\put(14.5,13.7){\line(1,0){54}}
\put(14.5,3.8){\line(1,0){54}}
\put(68.4,8.92){\oval(5,10.1)[rb]}
\put(68.4,6.6){\oval(5,10.1)[rt]}

\put(68.5,6.67){\oval(10,10.1)[rb]}
\put(68.5,8.68){\oval(10,10)[rt]}
\put(73.5,6){\line(0,1){4}}
\put(73.6,6){\vector(0,1){3}}
\put(70.9,6.0){\vector(0,1){2}}
\color{dunkelblau}

\end{picture}
\vspace{3mm}\caption{Layout of the optical fibre - coaxial cable detector, with L=5.0m.   10MHz RF signals come from the Rubidium atomic clock (Rb).  The Electrical to Optical converters (EO) use the RF signals to modulate 1.3$\mu$m infrared signals that propagate through the single-mode optical fibres.  The Optical to Electrical converters (OE) demodulate that signal and give the two RF signals that finally reach the Digital Storage Oscilloscope (DSO), which measures their phase difference.   The key effects are that the propagation speeds through the coaxial cables and optical fibres respond differently to their absolute motion through space, with no Fresnel drag in the coaxial cables, and Fresnel drag effect in the optical fibres. Without this key difference this detector does not work.    \label{fig:FUGWD}}
\end{figure}

An earlier 1st order in $v/c$ gravitational wave detector design is shown in Fig.\ref{fig:FUGWD}, with some data shown in Fig.\ref{fig:STEffect}. Only now is it known why that detector also worked, namely that there is a Fresnel drag effect in the optical fibres, but not in the RF coaxial cable. Then the travel time difference, measured at the DSO, is given by 
\vspace{-2mm} \begin{equation}
\Delta t=\frac{2v\cos(\theta)L(n_1^2-1)}{c^2}+..
\label{eqn:OFRF}\end{equation}
where $n_1$ is the effective refractive index of the RF coaxial cable.
Again the data is in remarkable agreement with the flyby determined ${\bf v}$.

\section{2nd Order in $v/c$ Gas-Mode Detectors}
\begin{figure}[h]
\vspace{-0mm}
\hspace{5mm}\includegraphics[scale=0.8]{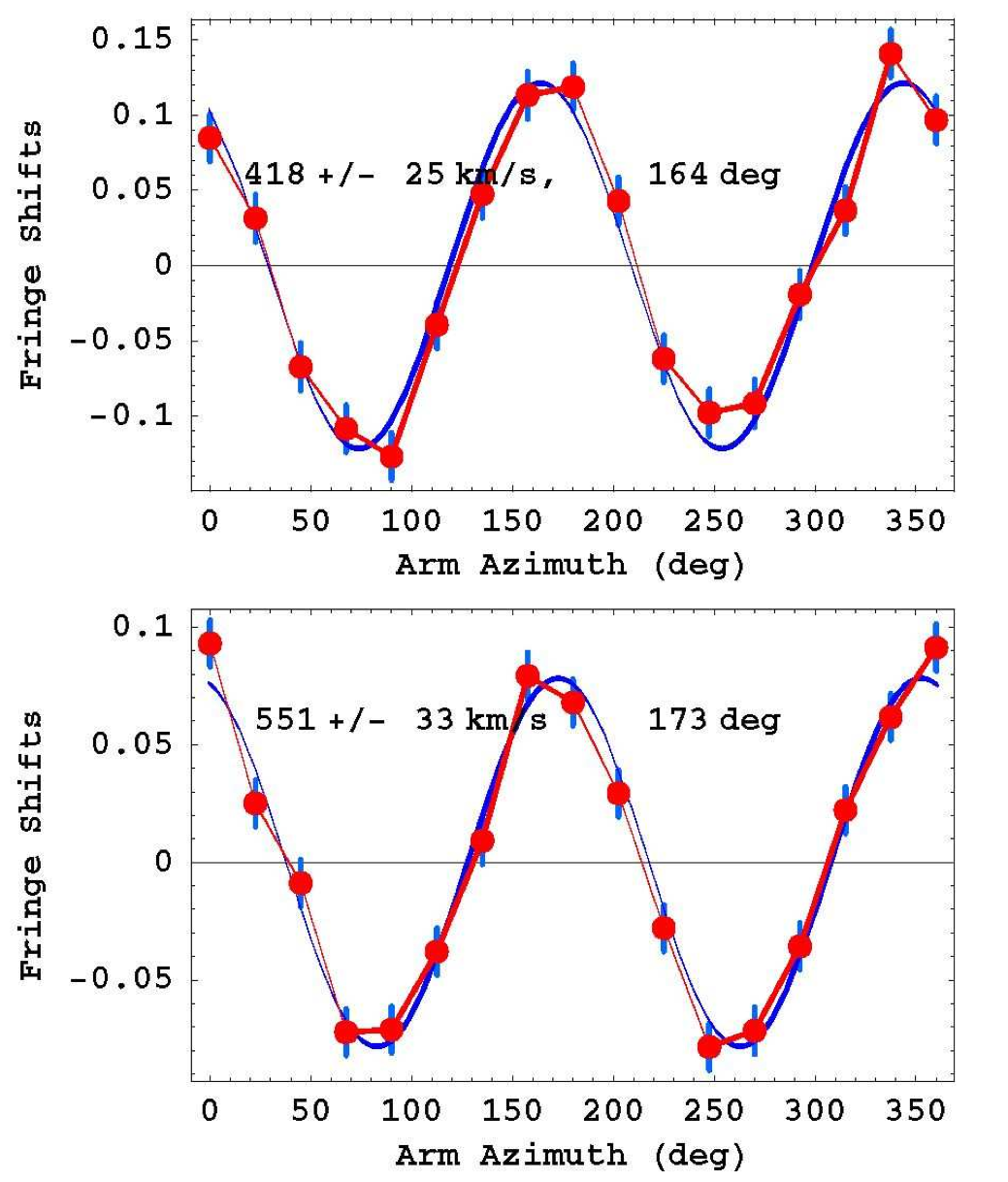}
\caption{\small {Top: Typical Miller data from 1925/26 gas-mode Michelson interferometer, from 360$^\circ$ rotation.  Bottom: Data from Michelson-Morley 1887 gas-mode interferometer, from 360$^\circ$ rotation.}}
\label{fig:MandMM}\end{figure}

\begin{figure}
\vspace{0mm}
\hspace{5mm}\includegraphics[scale=0.4]{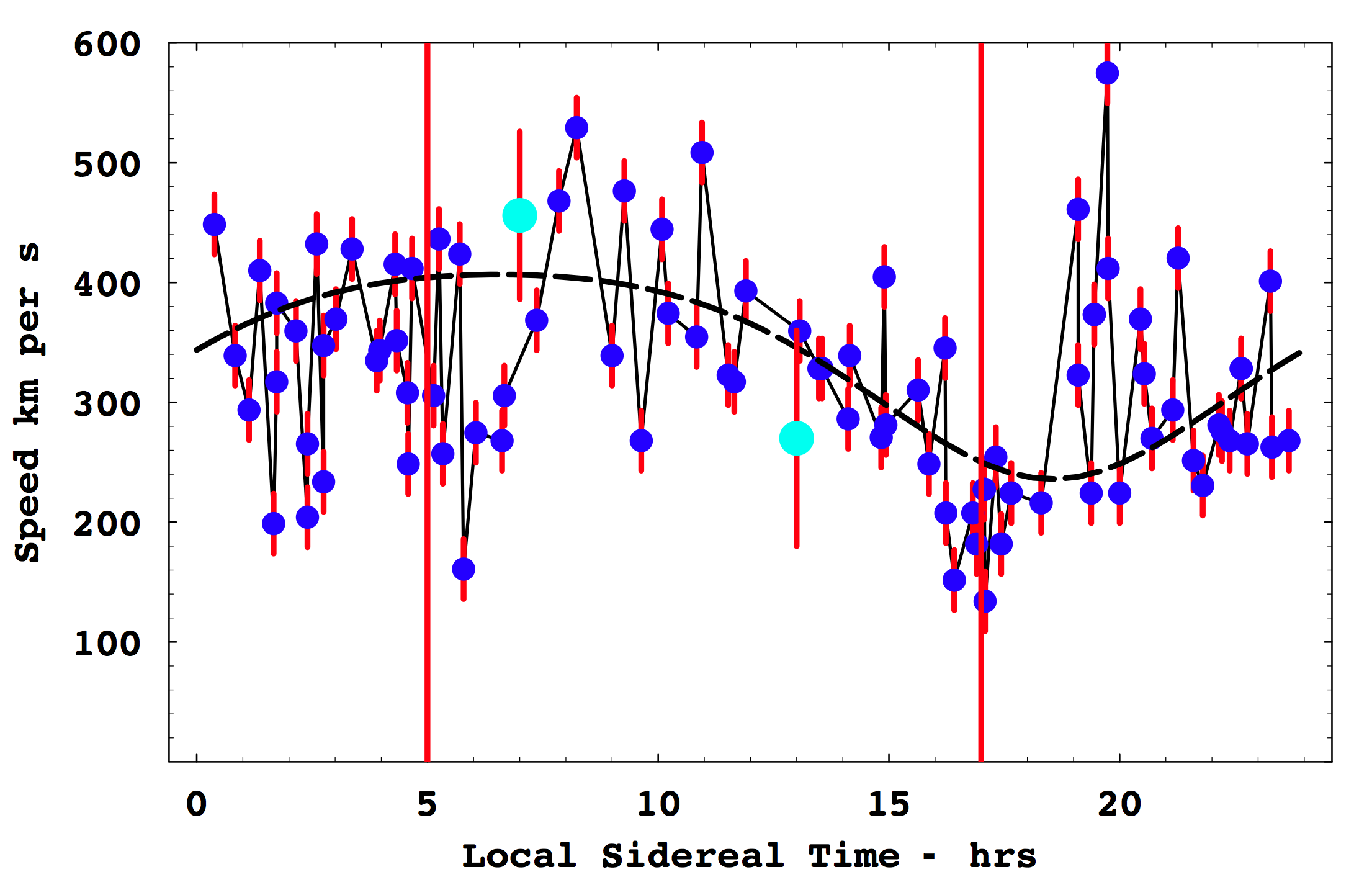}
\caption{\small {Miller data for composite day in September 1925, and also showing Michelson-Morley 1887 July data at local sidereal times of 7$^h$ and 13$^h$. The waved/turbulence effects are very evident, and comparable to data reported herein from the new detector.}}
\label{fig:Miller}\end{figure}

It is important that the gas-mode 2nd order in $v/c$ data from Michelson and Morley, 1887, and from Miller, 1925/26, be reviewed in the light of the recent experiments and flyby data.
Shown in Fig.\ref{fig:MandMM} (top) is   Miller data    from September 16, 1925, $4^h 40^\prime$  Local Sidereal Time (LST) - an average of data from 20 turns of the gas-mode Michelson interferometer. Plot and data after fitting and then subtracting both the temperature drift and Hicks effects from both, leaving the expected sinusoidal form.  The error bars are determined as the rms error in this fitting procedure, and show how exceptionally small were the errors, and which agree with Miller's claim for the errors. Best result  from the Michelson-Morley 1887 data - an average of 6 turns, at  $7^h$  LST on July  11, 1887,  is shown in Fig.\ref{fig:MandMM} (bottom).  Again the rms error is remarkably small.  In both cases the indicated speed is  $v_P$ - the 3-space speed projected onto the plane of the interferometer. The angle is  the azimuth of the  3-space speed projection at the particular LST.  Fig.\ref{fig:Miller}  shows speed fluctuations from day to day significantly exceed these errors, and reveal the existence of 3-space flow turbulence - i.e gravitational waves.    The data shows considerable fluctuations, from hour to hour, and also day to day, as this is a composite day.  The dashed curve shows the non-fluctuating  best-fit sidereal effect variation over one day, as the earth rotates, causing the projection onto  the plane of the interferometer of the  velocity of the average direction of the space flow to change.  The  predicted projected sidereal speed variation for Mt Wilson is 251 to 415 km/s, using the Casinni flyby data and   the STP air refractive index of $n=1.00026$  appropriate atop Mt. Wilson, and the  min/max occur at approximately 5hrs and 17hrs local sidereal time (Right Ascension).  For the Michelson-Morley experiment in Cleveland the predicted projected sidereal speed variation is 239 to 465 km/s. Note that the Cassini flyby in August gives a RA$= 5.15^h$, close to the RA apparent in the above plot.  The green data points, showing daily fluctuation  bars, at $7^h$ and $13^h$, are from the Michelson-Morley  1887 data, from averaging (excluding only the July 8 data for 7$^h$ because it has poor S/N), and with same rms error analysis. The fiducial time lines are at $5^h$ and $17^h$. The data indicates the presence of  turbulence in the 3-space flow, i.e gravitational waves, being first seen in the Michelson-Morley experiment. 

\section{Conclusions}
The Dual RF Coaxial Cable Detector exploits the Fresnel drag anomaly in RF coaxial cables, {\it viz}  the drag effect is absent in such cables, for reasons unknown, and this 1st order in $v/c$ detector is compact, robust and uses one clock.   This anomaly now explains the operation of the Optical-Fibre - Coaxial Cable Detector, and permits a new calibration.  These detectors have confirmed the absolute motion of the solar system  and the gravitational wave effects seen in the earlier experiments of Michelson-Morley, Miller, DeWitte and Torr and Kolen. Most significantly these experiments agree with one another, and with the absolute motion velocity vector determined from spacecraft earth-flyby Doppler shifts. The observed significant wave/ turbulence effects reveal that the so-called ``gravitational waves" are easily  detectable in small-scale laboratory  detectors, and are considerably larger than those predicted by GR. These effects are not detectable  in vacuum-mode Michelson terrestrial interferometers, nor by their analogue vacuum-mode resonant cavity experiments.   

The new  Dual RF Coaxial Cable Detector permits a detailed study and characterisation of the wave effects, and with the detector having the inclination equal to the local latitude the earth rotation effect may be removed, as the detector is then parallel to the earth's spin axis,  enabling a more accurate characterisation of  the wave effects. The major discovery arising from these various results is that 3-space is directly detectable and has a  fractal textured structure. This and numerous other effects are consistent with the dynamical theory for this 3-space. We are seeing the emergence of fundamentally new physics, with space being a a non-geometrical dynamical system, and fractal down to the smallest scales describable by a classical velocity field, and below that by  quantum foam dynamics \cite{Book}. Imperfect and incomplete is the geometrical model of space.

\section{Acknowledgements}
The Dual RF Coaxial Cable Detector is part of the Flinders University Gravitational Wave Detector Project.  The DSO, Rb RF frequency source and coaxial cables  were funded by an Australian Research Council Discovery Grant: {\it Development and Study of a New Theory of Gravity}.  Special thanks to CERN for donating  the phase stabilised optical fibre, and to  Fiber-Span for donating the optical-electrical converters. Thanks for support to Professor Warren Lawrance, Bill Drury, Professor Igor Bray, Finn Stokes and Dr David Brotherton.

\small{

}


\begin{thebibliography}{99}

  \bibitem{CahillGW} Cahill R.T. {\it A New Light-Speed Anisotropy Experiment: Absolute Motion and Gravitational Waves},   {\it Progress in Physics},  v.  4, 73-92, 2006.  
\bibitem{CahillBrisbane}  Cahill R.T. {\it  Absolute Motion and Gravitational Wave Experiment Results}, Contribution to {\it Australian Institute of Physics National Congress}, Brisbane, Paper No. 202, 2006.


\bibitem{MMCK} Cahill R.T. and Kitto K. {\it Michelson-Morley Experiments Revisited}, {\it Apeiron}, v. 10(2), 104-117, 2003.
 \bibitem{MMC}  Cahill  R.T. {\it The Michelson and Morley 1887 Experiment
and the Discovery of Absolute Motion},   {\it Progress in Physics},  v.  3, 25-29, 2005.
\bibitem{Miller}   Miller D.C. {\it  The Ether-Drift Experiment and the Determination of the Absolute Motion of the Earth}, {\it Rev. Mod. Phys.},  v. 5, 203-242, 1933.
\bibitem{Torr}  Torr D.G. and Kolen P. {\it An Experiment to Measure Relative Variations in the One-Way Velocity of Light}, in  {\it Precision Measurements and Fundamental Constants II},  Taylor B.N. and  Phillips W.D.  eds. {\it  Natl. Bur. Stand. (U.S.), Spec. Pub.}, 617,  675-679, 1984.
\bibitem{DeWitte}  Cahill R.T. {\it The Roland De Witte 1991 Experiment}, {\it Progress in Physics}, v.  3, 60-65, 2006.

\bibitem{And2008} Anderson J.D., Campbell J.K., Ekelund J.E., Ellis J. and Jordan J.F. {\it  Anomalous Orbital-Energy Changes Observed during Spacecraft Flybys of Earth},  {\it Phys. Rev. Lett.}, v. 100,  091102, 2008.

  \bibitem{CahillNASA} Cahill R.T. {\it Combining NASA/JPL One-Way Optical-fibre   Light-Speed  Data with  Spacecraft Earth-Flyby Doppler-Shift Data  to Characterise  3-Space Flow},   {\it Progress in Physics},  v. 4, 50-64, 2009.  
  


\bibitem{CahillMink}  Cahill R.T. {\it Unravelling Lorentz Covariance and the Spacetime Formalism}, {\it Progress in Physics},  v. 4, 19-24, 2008.

  \bibitem{CahillOF} Cahill R.T.  and Stokes F. {\it  Correlated Detection of sub-mHz Gravitational Waves by Two Optical-fibre Interferometers},   {\it Progress in Physics}, v. 2, 103-110, 2008.
  
  
  \bibitem{Book} Cahill  R.T. {\it Process Physics: From Information Theory to Quantum Space and Matter},  {\it Nova Science Pub.}, New York, 2005.   
  
\bibitem{CahillFresnel}  Cahill R.T. and Brotherton D.,   {\it Experimental Investigation of the Fresnel Drag Effect in RF Coaxial Cables}, {\it Progress in Physics},  v. 1,  43-48, 2011.
    

     





\end{thebibliography}
\end{document}